\documentclass[runningheads]{llncs}
\usepackage[T1]{fontenc}
\usepackage{graphicx}
\usepackage{acronym}
\usepackage{adjustbox}
\usepackage{tablefootnote}
\usepackage{bm}
\usepackage{caption}
\usepackage[labelformat=simple]{subcaption}
\usepackage{graphicx}
\usepackage{amsmath}
\usepackage{booktabs}
\usepackage{threeparttable}
\usepackage{multirow}
\usepackage{enumitem}
\usepackage{xcolor}
\usepackage{mathtools}
\usepackage{marvosym}
\usepackage[ruled]{algorithm2e}
\usepackage{tablefootnote}
\usepackage{soul, color, xcolor}
\usepackage{amssymb}
\usepackage{pifont}
\usepackage{wasysym}
\usepackage{makecell}
\usepackage{hyperref}

\begin{document}
\title{Reproducibility Analysis and Enhancements for Multi-Aspect Dense Retriever with Aspect Learning}

%
\titlerunning{Reproducibility Analysis and Enhancements}
%
%
%

\author{Keping Bi\inst{1,2}\orcidID{0000-0001-5123-4999} \and
Xiaojie Sun\inst{1,2}\orcidID{0009-0006-4570-6359} \and \\
Jiafeng Guo\inst{1, 2}\thanks{Corresponding author}\orcidID{0000-0002-9509-8674} \and 
Xueqi Cheng\inst{1,2}\orcidID{0000-0002-5201-8195}}

\authorrunning{K. Bi et al.}
%
\institute{CAS Key Lab of Network Data Science and Technology, ICT, CAS \and
University of Chinese Academy of Sciences \\
\email{\{bikeping, sunxiaojie21s, guojiafeng, cxq\}@ict.ac.cn}
 }
 
\maketitle              
\begin{abstract}
Multi-aspect dense retrieval aims to incorporate aspect information (e.g., brand and category) into dual encoders to facilitate relevance matching. 
As an early and representative multi-aspect dense retriever, MADRAL learns several extra aspect embeddings and fuses the explicit aspects with an implicit aspect ``OTHER'' for final representation. 
MADRAL was evaluated on proprietary data and its code was not released, making it challenging to validate its effectiveness on other datasets. We failed to reproduce its effectiveness on the public MA-Amazon data, motivating us to probe the reasons and re-examine its components. We propose several component alternatives for comparisons, including replacing ``OTHER'' with ``CLS'' and representing aspects with the first several content tokens. Through extensive experiments, we confirm that learning ``OTHER'' from scratch in aspect fusion is harmful. In contrast, our proposed variants can greatly enhance the retrieval performance. 
Our research not only sheds light on the limitations of MADRAL but also provides valuable insights for future studies on more powerful multi-aspect dense retrieval models. Code will be released at: \url{https://github.com/sunxiaojie99/Reproducibility-for-MADRAL}.



\end{abstract}
%



%
%

\vspace*{-2mm}
\section{Introduction}
Standing on the shoulders of pre-trained language models (PLMs)\cite{bert,ernie}, dense retrieval models have exhibited impressive performance in the first stage of information retrieval \cite{condenser,co-condenser,costa,bibert-pretrain}. Most dense retrieval models concentrate on unstructured textual data, while much less attention has been paid to structured item retrieval such as product search and people search. These scenarios have a wide population of users and the aspect information like brand (e.g., ``Apple") and affiliation (e.g., ``Microsoft'') can be pivotal to enhance relevance matching. Nonetheless, it remains largely unexplored how to effectively integrate these aspects within dense retrieval models.

\begin{figure}[t]
\setlength{\belowcaptionskip}{-0.1cm}
\setlength{\abovecaptionskip}{-0.3cm}
    \centering
    \setlength{\abovecaptionskip}{0.1cm}
    \includegraphics[scale=0.4]{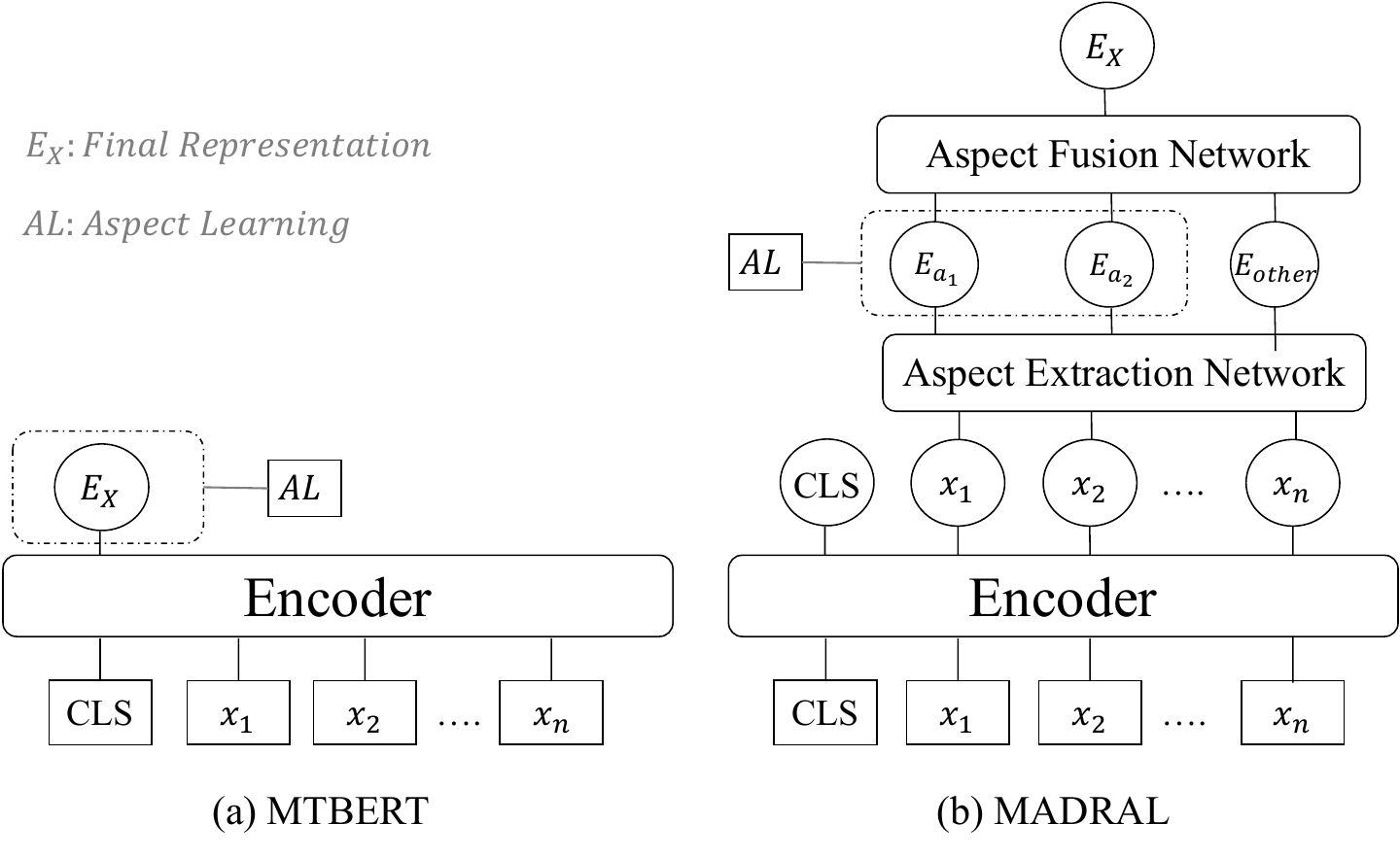}
    \caption{Two multi-aspect dense retrieval models proposed by Kong et al. \cite{madr}.}
    \label{fig:mtbertmadralmodel}
\vspace{-2mm}
\end{figure}

Recently, Kong et al. \cite{madr} initiated such a study by proposing a Multi-Aspect Dense Retriever with Aspect Learning, named MADRAL, and a simpler yet competitive baseline MTBERT.
As illustrated in Figure \ref{fig:mtbertmadralmodel}, MADRAL has three major components, i.e., aspect extraction, aspect learning, and aspect fusion, to produce the final representation $E_X$.
Specifically, this model employs an aspect extraction network to extract extra aspect embeddings alongside the initial BERT parameters and conducts aspect learning by predicting the value IDs of an aspect (e.g., the ID of ``Beauty'' in the vocabulary of the product category). 
Notably, a special aspect ``OTHER'' is included to capture the implicit semantics that the explicit aspects cannot cover. 
Then for relevance matching, these aspect embeddings are integrated using an aspect fusion network to produce the final query/item representation.
In contrast, MTBERT only conducts aspect learning on the CLS token, which is also used for relevance matching.
Both models significantly outperform the original BERT and MADRAL can achieve much more compelling performance. The framework of MADRAL is insightful for the research on multi-aspect dense retrieval.   

Although claimed to be effective, MADRAL has been experimented on proprietary data (i.e., Google shopping) that is not accessible to the public. The code of MADRAL has not been released either, which makes it even harder to reproduce the experimental results in \cite{madr}. Since Google shopping data has aspect information of both queries and items, it is also unknown whether MADRAL will be effective on other datasets of different properties. We have tried to reproduce its performance on the public MA-Amazon data, which has large-scale real-world queries and multiple aspect information associated with the items, but surprisingly find that MADRAL \footnote{The authors have not provided their code upon our request but verified our implementation of MADRAL.} has significantly worse performance than its backbone BERT. This has motivated us to study why it does not work and how to enhance it to work effectively.

We speculate that there are two potential reasons for its unsatisfactory performance: 1) the brand-new embedding of aspect ``OTHER'' may not learn the implicit semantics well during fine-tuning; 2) it is challenging to learn extra aspect embeddings sufficiently from scratch during pre-training. To validate the reasons, we propose several alternative methods for aspect fusion and aspect representation. Specifically, instead of learning implicit semantics with a new token ``OTHER'', we propose to fuse ``CLS'', which is designated to capture global content semantics explicitly, in the final representation. For aspect representation, we reuse the first several content tokens to represent aspects whose embeddings only need to be adjusted with the aspect learning objectives. Extensive experiments show that both versions of enhancements can yield significantly better retrieval results when replacing the original counterparts of MADRAL, confirming the existence of the above issues. Our studies pave the way for future research on this topic that uses MADRAL as a benchmark and also provide valuable insights into the development of more powerful multi-aspect dense retrievers. 

\vspace*{-2mm}
\section{RELATED WORK}

\vspace*{-2mm}
\textbf{Dense Retrieval}. Dense retrieval models typically adopt a bi-encoder architecture, which encodes a query and a document into two vectors and uses a similarity function like a dot product to measure their relevance. Karpukhin et al. \cite{DPR} have explored pre-trained language models (PLMs) for information retrieval by using the BERT as the encoder and training it with in-batch negatives. This achieves superior performance compared to the models before the PLM era. Subsequently, researchers delved into various fine-tuning techniques to improve dense retrieval such as mining hard negatives \cite{ance,rocketqa}, distilling the knowledge from cross encoders \cite{distil-cross}, and representing documents with multi-vector representations \cite{me-bert,col-bert,mvr}. 

Most research efforts on dense retrieval have been spent on unstructured text until recently Kong et al. \cite{madr} proposed an effective method MADRAL that incorporates the structured aspect information of queries or items into the dense retrievers. This work leverages a typical way of injecting the aspect information \cite{ai2018learning,rerank-explain} to the item representation, i.e., predicting the values associated with the aspects as an auxiliary training objective. 
Following this paradigm, Sun et al. \cite{sun2023multigranularityaware} studied how to capture fine-grained semantic relations between different aspect values.
As MADRAL \cite{madr} is the first multi-aspect dense retriever and adopts a typical manner of modeling the aspects, our reproducibility study on it will pave the way for future research that uses MADRAL as a benchmark in this direction.

\noindent
\textbf{Multi-Field Retrieval}.
It has been a longstanding research topic on how to effectively utilize multi-field information such as titles, keywords, and descriptions in documents. The earliest attempt can date back to BM25F \cite{robertson2004simple}. More recently, Liu et al. \cite{field-liu} explored the incorporation of multi-field information into the relevance models.
Prior to the advent of PLMs, researchers have investigated leveraging the document fields in neural ranking models \cite{field-Balaneshinkordan,field-Choi,field-Zamani}. For example, Zamani et al. \cite{field-Zamani} proposed to aggregate field-level representations using a matching network and trained the model with field-level dropout. 
There has been ongoing research on the utilization of multi-field information \cite{field-biomedical,field-shan2023beyond} after PLMs have become the dominant retriever backbone. For instance, Shan et al. \cite{field-shan2023beyond} proposed to leverage field-level cross interactions between queries and items as an auxiliary fine-tuning objective to improve retrieval performance. Sun et al. \cite{sun2023pre} treated item aspects as text strings and proposed a pre-training method to enhance the retriever. 

Although the aspects can be simply treated as fields, multi-field retrievers only focus on the document side and cannot handle the case that query aspects are also available. Moreover, aspects and fields have some essential differences: fields are comprised of unstructured text that has infinite semantic space, whereas an aspect is defined by a finite set of values, serving as the aspect annotations. Consequently, multi-aspect and multi-field retrieval face distinct challenges. 
MADRAL \cite{madr} has been experimented on the data having aspects for both queries and items, and it was not compared to any baselines that treat aspects as fields. Since we use the public MA-Amazon dataset that only has item aspects, we also include a straightforward baseline that uses aspects as fields and concatenates the aspect texts, i.e., BIBERT-CONCAT in Section \ref{sucsec:exp_methods}. 



\vspace*{-4mm}
\section{Preliminaries of Multi-Aspect Dense Retrievers}
\label{sec:preliminary}

\vspace*{-2mm}
\textbf{Task Definition}. In multi-aspect dense retrieval, queries and candidate items can have multiple aspects such as brand, color, and category. Given a query $q$ or item $i$, each of its associated aspects $a$ has a finite vocabulary of value set, denoted as $V_a$, and an embedding lookup table $T_a \in \mathbb{R}^{|V_a| \times H}$, where each value ID maps to an $H$-dimensional vector. Suppose that the aspect set is $A$ containing $k$ aspects, i.e., $A = {a_1, a_2, \cdots, a_k}$, their corresponding annotated value sets are $\mathcal{A}_{a_1}, \mathcal{A}_{a_2}, \cdots, \mathcal{A}_{a_k}$. The content tokens of $q$ or $i$ (that can include titles and descriptions) are denoted as $X = x_1,x_2,\cdots,x_n$. A multi-aspect dense retrieval model aims to learn effective representations of $q$ and $i$ by incorporating the aspect information and capturing the content semantics so that their similarities can reflect their relevance. 

\begin{figure}[t]
\setlength{\belowcaptionskip}{-0.1cm}
\setlength{\abovecaptionskip}{0cm}
    \centering
    \setlength{\abovecaptionskip}{0.1cm}
    \includegraphics[scale=0.3]{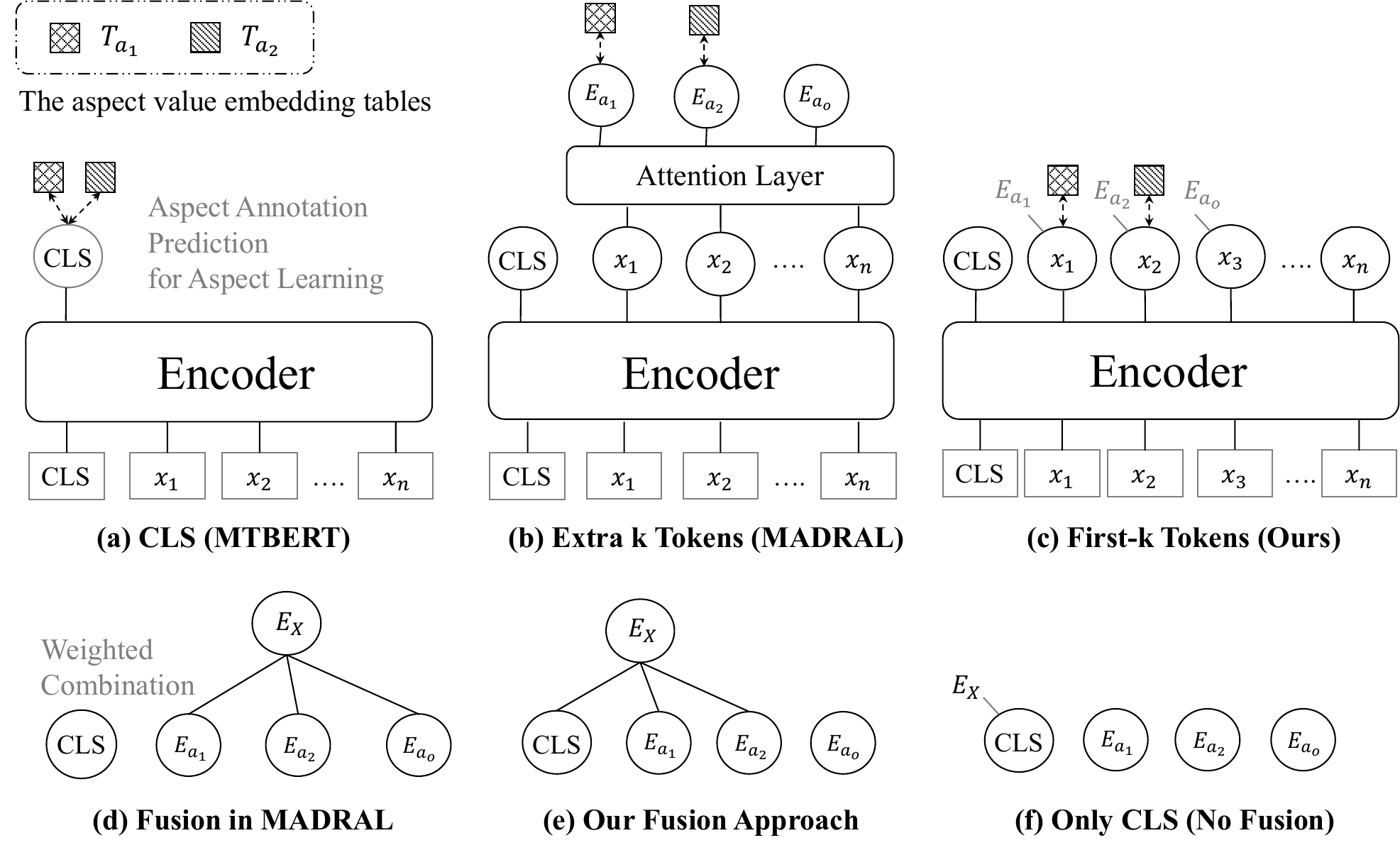}
    \caption{The upper figures illustrate the aspect representation and learning of MTBERT, MADRAL, and our variant. The lower figures show aspect fusion methods to yield the final representation $E_X$. The weighted combination can be CLS gating or presence weighting. $a_o$ denotes the special aspect ``OTHER''.}
    \label{fig:model}
\vspace*{-3mm}
\end{figure}

\noindent
\textbf{Multi-aspect Dense Retrievers with Aspect Learning}. 
As shown in Figure \ref{fig:mtbertmadralmodel} and \ref{fig:model}, typical multi-aspect dense retrievers \cite{madr} usually have three major components: 1) \textit{Aspect Representation}, that either declares extra aspect embeddings (e.g., in MADRAL) or reuses the ``CLS'' token (e.g., in MTBERT) to capture the aspect information; 2) \textit{Aspect Learning}, that injects the aspect-value information into the aspect representation by predicting its associated value IDs during pre-training (may also be beneficial in fine-tuning); 3) \textit{Aspect Fusion} that merges the learned aspect representations into the final query/item representation for relevance matching during fine-tuning. In MTBERT, all the aspect learning is conducted on the ``CLS'' token, so no additional fusion is needed. 

In the next two sections, we elaborate on the component variants of aspect representation and aspect fusion we study. Since queries and items have the same learning process, we only use items in the illustration for brevity. 

\vspace*{-2mm}
\section{Component Variants: Aspect Representation}
\label{sec:aspect_rep}
\vspace*{-3mm}
To effectively incorporate the aspect-value information into an item, reasonable aspect representation is pivotal. Figure \ref{fig:model}(a), \ref{fig:model}(b), and \ref{fig:model}(c) show the three variants of aspect representations we compare. 

\vspace*{-2mm}
\subsubsection{Reusing CLS Token (MTBERT).}
\label{subsec:reuse_cls}
MTBERT \cite{madr} reuses the ``CLS'' token to conduct aspect learning (see Figure \ref{fig:model}(a)). It naturally injects aspect annotations of an item into the ``CLS'' token that also captures content semantics. The aspect information can be learned on top of ``CLS'', which can be a decent starting point. 
However, it compresses the information from multiple aspects and the content into a single token without weighting mechanisms. So, it does not differentiate the importance of each information source to relevance ranking, which could yield suboptimal retrieval results.

\vspace*{-5mm}
\subsubsection{Declaring k Extra Embeddings (MADRAL).}
\label{subsec:extra_k}
MADRAL \cite{madr} represents aspects with extra embeddings (Figure \ref{fig:model}(b)), that are computed based on the attention over the encoded content tokens $Encoder(X)$ $=Encoder(x_1,x_2,\cdots,x_n)$, where $Encoder$ can be any transformer-based encoders like BERT. The aspect embeddings $E_A$, stacked from ${E_{a_1}, E_{a_2},\cdots, E_{a_k}}$, is computed as follows:
\begin{equation}
\setlength{\abovedisplayskip}{1pt}
\setlength{\belowdisplayskip}{1pt}
    \begin{aligned}
    & E_A = Attention(Q W^Q, Encoder(X) W^K, Encoder(X) W^V), \\
    & Attention(Q,K,V) = softmax(\frac{QK^T}{\sqrt{H}})V.
    \end{aligned}
\end{equation}
Attention is the multi-head attention function involving $Q,K,V$ in the standard transformer \cite{vaswani2017attention}. $Q$ is the set of aspect embeddings in this case. In this way, each aspect has its own representation and can be dedicated to its own learning. The influence of each aspect on the final representation can be automatically learned.
However, in MADRAL, these new parameters are only learned from the aspect learning objectives introduced in Section \ref{sec:al}, which could be challenging to learn them well from scratch, especially when there are not many aspect annotations. 

\vspace*{-5mm}
\subsubsection{Reusing First-k Content Tokens.}
\label{subsec:reuse_first_k}
Instead of training extra k aspect embeddings from scratch, we propose an alternative approach that reuses the encoder output of the first k content tokens, i.e., $x_1, x_2, \cdots, x_k$, to represent the k-associated aspects (shown in Figure \ref{fig:model}(c)).
In MADRAL, the embeddings of content tokens are loaded from a pre-trained BERT and also updated by the masked language model (MLM) loss when ingesting the local corpus. Hence, they are learned more sufficiently and can serve as better starting points than brand-new extra tokens. The tokens at the beginning of the content are usually important to represent the content semantics. Guiding these tokens with the aspect learning loss is a way that not only binds the aspect information to content tokens like MTBERT does but also differentiates the influence of each aspect on the final representation like MADRAL does. So, it can leverage the advantages from both perspectives. 


\vspace*{-4mm}
\section{Component Variants: Aspect Fusion}
\label{sec:aspect_fusion}
During relevance matching, the aspect embeddings are fused into a single item embedding $E_X$, i.e., 
\begin{equation}
\setlength{\abovedisplayskip}{1pt}
\setlength{\belowdisplayskip}{1pt}
\label{eq:fusion}
    E_X = \sum_{a\in A} w_a E_a.
\end{equation}
MTBERT does not have the phase of aspect fusion and uses the CLS token directly for relevance matching. MADRAL \cite{madr} has three fusion networks: weighted sum, CLS-gating, and presence weighting. 
Since the first one does not perform well \cite{madr}, we only study the latter two. We will elaborate from two perspectives: the weighting mechanism and the objects to fuse. 

\vspace*{-2mm}
\subsection{Weighting Mechanism}
\label{subsec:weighting}
\noindent%
\textbf{CLS Gating.}
The encoded embedding of ``CLS'' is projected to $|A|$ (i.e., the number of aspects) logits, with a linear layer: $Linear(E_{CLS})\in \mathbb{R}^{|A|}$, and the softmax weight computed for each logit is used as the final weight. In other words, taking one logit $\gamma_a \in Linear(E_{CLS})$ for example, $w_a = \frac{e^{\gamma_a}}{\sum_{a'\in A}e^{\gamma_{a'}}}$.  

\noindent%
\textbf{Presence Weighting.}
This mechanism computes the weight of each aspect according to its presence probability in the item, i.e., $w_a = P(I_a) \gamma_a$, where $I_a$ indicates that item $I$ has annotated values for $a$, $P(I_a) = Sigmoid(E_a)$, and $\gamma_a$ is a learned parameter. $P(I_a)$ is learned with a cross-entropy loss based on whether an aspect has associated values in an item, which will be described in Section \ref{subsec:app}. 

\vspace*{-3mm}
\subsection{Objects to Fuse}
\label{subsec:obj_fusion}

\vspace*{-2mm}
The objects to fuse into the final item representation have a huge impact on retrieval performance, which we will show in the experiments. For both CLS-Gating and presence weighting, besides the original objects MADRAL fuses, we propose an alternative approach for fusion that slightly revises the fusion objects and can greatly enhance the performance. We introduce both ways as follows: 

\noindent%
\textbf{Aspect and Implicit Token (``OTHER'').}
In MADRAL, for both CLS-Gating and Presence Weighting, besides the standard aspects like brand and color, it adds a special aspect ``OTHER'' to capture the important information that may not be included in the explicit aspects. No aspect learning is conducted on this special aspect and it is supposed to learn implicit semantics automatically. If we denote the special aspect ``OTHER'' as $a_o$, the $k$ elements in the set $A$ in Equation \eqref{eq:fusion} becomes: 
\begin{equation}
\setlength{\abovedisplayskip}{1pt}
\setlength{\belowdisplayskip}{1pt}
\label{eq:fuse_ao}
A = \{{a_1}, {a_2}, \cdots, {a_{k-1}}, {a_o}\}.
\end{equation}
The fusion weights and the embedding of $a_o$ need to be learned during fine-tuning. When there is not sufficient training data with relevance labels to fine-tune the retriever, it could be difficult for the model to learn them well, especially $a_o$, which inevitably harms the retrieval performance.



\noindent%
\textbf{Aspect and Explicit Token (``CLS'').}
We believe that an effective item representation should capture both content semantics and the aspect information of the item.
Rather than fusing with the embedding of ``OTHER'' that learns implicit information, we use the ``CLS'' token that captures the global item semantics explicitly as a pseudo aspect in the fusion. In other words, the set $A$ in Equation \eqref{eq:fusion} becomes:
\begin{equation}
\setlength{\abovedisplayskip}{1pt}
\setlength{\belowdisplayskip}{1pt}
    \label{eq:fusion_a_c}
    A = \{{a_1}, {a_2}, \cdots, {a_{k-1}}, {CLS}\}.
\end{equation}
Then, only fusion weights need to be learned during fine-tuning and the objective is clear: balancing the effects of content and aspects on the final representation for relevance matching. 

\noindent%
\textbf{Only CLS (No Aspect Fusion).}
Though we have pre-trained the aspect embeddings, during relevance matching (or fine-tuning), we do not fuse these aspect embeddings but instead use the embedding of ``CLS'' as the item embedding. 
In this way, the aspect learning process conducted on the extra k embeddings can be considered as purely multi-task learning that could guide the underlying parameters in the encoder to a better optimum. Then, the content tokens also carry some aspect information and the final ``CLS'' embedding could be a better representation for relevance matching.

\vspace*{-3mm}
\section{Aspect Learning and Overall Training Objectives}
\label{sec:al}
\vspace*{-2mm}
\noindent
\textbf{Aspect Prediction (AP).}
\label{subsec:ap}
The typical way of learning aspect embeddings is to predict the annotated value IDs of the aspects \cite{ai2018learning,rerank-explain}. MADRAL\cite{madr} also adopts this method. Take an arbitrary aspect $a$ for instance, given its  
ground-truth annotation set $\mathcal{A}_a$ and its global value set $V_a$, the loss function is:   
\begin{equation}
\setlength{\abovedisplayskip}{1pt}
\setlength{\belowdisplayskip}{1pt}
\label{eq:ap}
    \mathcal{L}_{AP}^a = - \sum_{v^+\in \mathcal{A}_a} log \frac{exp(E_a \cdot E_{v^{+}})}{\sum_{v \in V_a}{exp(E_a \cdot E_v)}},
\end{equation}
where $E_v$/$E_{v^+} \in \mathbb{R}^H$ is the aspect value embedding from $a$'s embedding lookup table $T_a$. This is the major loss function to learn aspect embeddings. 

\noindent
\textbf{Aspect Presence Prediction (APP).}
\label{subsec:app}
MADRAL\cite{madr} also proposes the loss of predicting whether an item has a valid value for a certain aspect as a part of presence weighting (introduced in Section \ref{subsec:weighting}). The APP loss is essentially a binary classification loss:
\begin{equation}
\setlength{\abovedisplayskip}{1pt}
\setlength{\belowdisplayskip}{1pt}
\label{eq:app}
    \mathcal{L}_{APP}^a = -y_a \log(P(I_a)) - (1-y_a) \log (1-P(I_a)),
\end{equation}
where $y_a=0$ when $\mathcal{A}_a=\varnothing$ and $y_a=1$ otherwise.
Since this objective can also guide aspect embedding training even if not used in the weighting function, we include it as an aspect learning objective, supplementary to the AP loss. 

\noindent
\textbf{Pre-training Objective}. The backbone model is pre-trained on the local corpus to adapt the model parameters and learn the aspect embeddings. The overall pre-training objectives consist of the masked language model (MLM) loss and the aspect learning objectives that can be scaled with $\lambda_p$, i.e.,
\begin{equation}
\setlength{\abovedisplayskip}{1pt}
\setlength{\belowdisplayskip}{1pt}
\label{eq:pretrain-obj}
\mathcal{L}_{pretrain} = \mathcal{L}_{MLM} + \lambda_p \sum_{a \in A \setminus \{\text{OTHER}\} } (\mathcal{L}_{AP}^a + \mathcal{L}_{APP}^a).
\end{equation}
Note that $\mathcal{L}_{APP}^a$ is optional and only takes effect when needed. 

\noindent
\textbf{Fine-tuning Objective.}
Similarly, the loss for fine-tuning has relevance loss, and the aspect learning loss (with optional $\mathcal{L}_{APP}^a$) controlled by $\lambda_f$, i.e., 
\begin{equation}
\setlength{\abovedisplayskip}{1pt}
\setlength{\belowdisplayskip}{1pt}
\label{eq:finetune-obj}
\mathcal{L}_{finetune} = \mathcal{L}_{REL} + \lambda_f \sum_{a \in A \setminus \{\text{OTHER}\} } (\mathcal{L}_{AP}^a + \mathcal{L}_{APP}^a),
\end{equation}
where $\mathcal{L}_{REL}$ is a standard softmax cross-entropy loss that uses the relevant items and in-batch negative samples for training as in \cite{dual-encoder}. 

\vspace*{-2mm}
\section{EXPERIMENTAL SETTINGS}
\vspace*{-2mm}
\subsubsection{Multi-aspect Amazon ESCI Dataset (MA-Amazon).}
The MA-Amazon dataset \cite{sun2023pre} has 
482k products with the aspects of ``brand'', ``color'', and ``category'' besides their titles and descriptions. Only items have aspect information. The coverage of brand, color, and category of levels 1-2-3-4 on the items are 94\%, 67\%, and 87\%-87\%-85\%-71\%, respectively. The relevance dataset for fine-tuning has 17k, 3.5k, and 8.9k real-world queries for training, validation, and testing, respectively. For each query, the relevance dataset provides an average of 20.1 items, each accompanied by relevance judgments - ``Exact'', ``Substitute'', ``Complement'', and ``Irrelevant''. As in \cite{amazon-data,sun2023pre}, we treat \textit{Exact} as positive instances and the other judgments as negative during training and for recall calculation. 
Although MA-Amazon does not have query aspects as the private Google shopping data does, it is public and has large-scale real-world queries with relevance judgments. We are unaware of other such public datasets, so we only conduct experiments on MA-Amazon. 


\vspace*{-4mm}
\subsubsection{Methods for Comparison.}
\label{sucsec:exp_methods}
We compare the MADRAL variants with various dense retrieval baselines, some incorporating aspect information and others not. Besides the baselines the original MADRAL compares, we also include a baseline that uses the aspects as text strings rather than conducting aspect classification for its learning. The baselines are:
\textit{BIBERT}: A typical bi-encoder retriever and the backbone of MADRAL, using BERT's \texttt{CLS} token for query and item encoding. It is pre-trained with $\mathcal{L}_{MLM}$ in Eq. \eqref{eq:pretrain-obj} and fine-tuned with $\mathcal{L}_{REL}$ in Eq. \eqref{eq:finetune-obj};
\textit{Condenser}: An advanced pre-trained model for textual dense retrieval. It enhances the \texttt{CLS} embedding during pre-training by connecting middle-layer tokens to the top layers;
\textit{BIBERT-CONCAT}: A straightforward method that concatenates the text strings of aspect annotations with item content, adds an indicator token before each aspect, and also uses $\mathcal{L}_{MLM}$ for pre-training and $\mathcal{L}_{REL}$ for fine-tuning;
\textit{MTBERT}: A multi-task model based on BIBERT proposed in \cite{madr}, reusing CLS for aspect prediction alongside MLM during pre-training.
The MADRAL variants are: 
\textit{MADRAL-ori}: The best original MADRAL model \cite{madr} introduced in Section \ref{subsec:extra_k} \& \ref{subsec:weighting}; 
\textit{MADRAL-en-v1}: Our first enhanced version of MADRAL-ori that only refines it with the best aspect fusion method in Section \ref{sec:aspect_fusion}. They only differ during fine-tuning;
\textit{MADRAL-en-v2}: The second enhanced version of MADRAL-ori that incorporates the change in version 1 and the best aspect representation in Section \ref{sec:aspect_rep}. 

\vspace*{-4mm}
\subsubsection{Evaluation Metrics.}
\label{subsec:eval_metrics}
We use recall (R) and normalized discounted cumulative gain (NDCG) as evaluation metrics. Specifically, we report R@100, R@500, NDCG@10, and NDCG@50. As in \cite{amazon-data,sun2023pre}, the gains of E, S, C, and I judgments are set to 1.0, 0.1, 0.01, and 0.0, respectively. We conducted two-tailed paired t-tests (p < 0.05) to check statistically significant differences.

\vspace*{-4mm}
\subsubsection{Implementation Details.}
Since MADRAL does not release code, we have implemented all the MADRAL variants and the baseline methods on our own to ensure consistent experimentation details and fair comparisons. 
\textbf{Pre-training.}
For all methods, we share the encoder for both queries and items to promote knowledge sharing. In particular, we pre-train the models on the products in MA-Amazon to acquire the shared encoder for subsequent fine-tuning.
In line with prior research \cite{Ma2021PROPPW,costa}, we initialize all BERT components using Google's public checkpoint and employ the Adam optimizer with the linear warm-up technique. The learning rate and pre-training epoch are set to 1e-4 and 20 respectively. We accommodate a maximum token length of 156 and employ MLM mask ratios of 0.15.
For the scaling coefficient of AP and APP objectives, i.e., $\lambda_p$ in Eq.\ref{eq:pretrain-obj}, we slightly tune it from 0 to 0.5. Based on the validation results, it is set to 0.1. 
\textbf{Fine-tuning.}
For both datasets, we fine-tune all the models for 20 epochs. Following the previous work \cite{DPR}, we include a hard negative sample for each query besides in-batch negatives. We use a learning rate of 5e-6 and a batch size of 64. The maximum token lengths are set to 32 for queries and 156 for items. $\lambda_f$ in Eq.\ref{eq:finetune-obj} is scanned from \{0,0.05,0.1,0.2,0.3\}, and the best value is 0 according to evaluation results. We conduct fine-tuning on the pre-trained model checkpoints every two epochs and select the best-performing one on the validation set.

\vspace*{-5mm}
\section{Experimental Results}
\label{sec:exp_result}

\begin{table}[t]
\setlength\tabcolsep{9pt}
\renewcommand{\arraystretch}{0.8}
\setlength{\abovecaptionskip}{0pt}
\caption{Comparisons between various retrievers. $\dag$, $\ddag$, and $\ast$ indicate significant improvements over BIBERT, BIBERT-CONCAT, and MTBERT, respectively. The best overall and baseline results are underlined and bold.}
\centering
\label{tab:main_exp}
  \begin{tabular}{lcccc}
    \toprule
    Method &R@100&R@500 & NDCG@10 & NDCG@50 \\ 
    \midrule
    BIBERT & 0.6075 & 0.7795 & 0.3148 & 0.3929 \\ 
    Condenser & 0.6091 & 0.7801 & 0.3191 & 0.3960 \\ 
    \midrule
    BIBERT-CONCAT & 0.6137 & 0.7814 & \underline{0.3223} & \underline{0.4005} \\ 
    MTBERT & \underline{0.6139}$^{\dag}$ & \underline{0.7849}$^{\dag \ddag}$ & 0.3183$^{\dag}$ & 0.3969$^{\dag}$ \\ 
    \midrule
    \midrule
    MADRAL-ori & 0.5016 & 0.7121 & 0.2086 & 0.2823 \\ 
    \midrule
    MADRAL-en-v1 & 0.6159$^{\dag}$ & 0.7892$^{\dag \ddag \ast}$ & 0.3220$^{\dag \ast}$ & 0.4003$^{\dag \ast}$ \\ 
    MADRAL-en-v2 & \textbf{0.6219}$^{\dag \ddag \ast}$ & \textbf{0.7922}$^{\dag \ddag \ast}$ & \textbf{0.3291}$^{\dag \ddag \ast}$ & \textbf{0.4076}$^{\dag \ddag \ast}$ \\
    \bottomrule
  \end{tabular}
\vspace*{-6mm}
\end{table}
\vspace*{-3mm}
\subsection{Comparisons between MADRAL Variants and Baselines}
\vspace*{-1mm}
The results of the baselines, the original MADRAL, and our two variants of enhanced MADRAL are shown in Table \ref{tab:main_exp}. Among the baselines, we find that better pre-trained models (i.e., Condenser) have better performance, and incorporating the aspect information (MTBERT and BIBERT-CONCAT) can boost the retrieval performance. Note that the method that considers aspects as text strings, i.e., BIBERT-CONCAT, achieves competitive performance compared to methods that use aspects as auxiliary training objectives. For instance, the performance of MTBERT is better than BIBERT-CONCAT at lower positions (R@500) but worse at higher positions (NDCG@10,50). 
This indicates that treating aspects as text strings is also an effective approach to leverage these aspects, as observed in \cite{sun2023pre,field-shan2023beyond}. 
Yet, careful learning strategies are required when concatenating the aspect strings with the original content, especially on the query side \cite{sun2023pre}. It is also promising to study how to combine the two ways of using aspects (i.e., as text strings and by conducting associated value ID prediction) \cite{sun2023multigranularityaware}. Further discussions on leveraging aspects as strings are beyond the scope of this paper and interested users can refer to \cite{sun2023pre,field-shan2023beyond,sun2023multigranularityaware} for more information.

When we compare the performance of the MADRAL variants, it is surprising that the best results the original MADRAL can achieve (i.e., with presence weighting) are still worse than the baselines by a large margin. We attribute this to the insufficient learning of the special aspect - ``OTHER'' during fine-tuning. In contrast, the best variants we propose to enhance MADRAL in terms of the objects to fuse and aspect representation (denoted as ``-en-v1'' and ``-en-v2'' respectively) can significantly boost the retrieval performance. MADRAL-en-v1 uses ``CLS'' instead of ``OTHER'' and performs significantly better than MTBERT regarding almost all the metrics. It is also better than BIBERT-CONCAT at lower positions and similar at higher positions. Besides replacing ``OTHER'' with ``CLS'', MADRAL-en-v2 uses the first k content tokens to represent the aspects instead of declaring extra k aspect tokens. These two changes together lead to significantly better performance than all the baselines, showing that the proposed variants are effective ways of enhancements. 


\vspace*{-2mm}
\subsection{Comparisons of Fusion Methods}

If we only modify the fusion methods of MADRAL, only fine-tuning will be affected and the changes are relatively small. We introduce our studies on this part before aspect representation variants that incur more changes. Table \ref{tab:exp_fusion} shows how the fusion objects affect the retrieval performance when equipped with different aspect learning objectives (AP only or AP plus APP) during pre-training and multiple weighting mechanisms during fine-tuning. We can see that for both CLS-gating and presence weighting, using the token ``CLS'' to explicitly capture content semantics can boost the performance over learning implicit semantics with the token ``OTHER''. This confirms our speculation that learning the embedding of ``OTHER'' sufficiently from scratch during fine-tuning can be challenging, which may be a less severe issue on the Google Shopping data in \cite{madr} than on the MA-Amazon data as it has more training data for fine-tuning. 

\begin{table}
\setlength{\belowcaptionskip}{-0.4cm}
\renewcommand{\arraystretch}{0.8}
\setlength{\abovecaptionskip}{0pt}
\caption{Comparisons of various aspect fusion methods. The best results are in bold. All the methods that use either explicit token ``CLS'' in fusion (see Section \ref{subsec:obj_fusion}) or do not conduct fusion (see Section \ref{subsec:weighting}) are significantly better than fusion with implicit token ``OTHER''. $\dag$ indicates significant improvements over the BIBERT.}
\centering
\label{tab:exp_fusion}
  \begin{tabular}{lcccccc}
    \toprule
    Method &R@100&R@500 & NDCG@10 & R@100 & R@500  & NDCG@10 \\
    \midrule
    Aspect Fusion & \multicolumn{3}{c}{Explicit Token(``CLS'')}&\multicolumn{3}{c}{Implicit Token(``OTHER'')} \\
     \cmidrule(lr){2-4} \cmidrule(lr){5-7}
CLS-Gating & 0.6090 & 0.7832$^{\dag}$ & 0.3158 & 0.4743 & 0.7002 & 0.1809 \\ 
APP+CLS-Gating & 0.6118$^{\dag}$ & 0.7836$^{\dag}$ & \textbf{0.3194}$^{\dag}$ & 0.4717 & 0.6973 & 0.1744 \\ 
PresenceWeighting & \textbf{0.6148}$^{\dag}$ & \textbf{0.7878}$^{\dag}$ & 0.3184$^{\dag}$ & 0.5016 & 0.7121 & 0.2086 \\ 
\midrule
NoAspectFusion & 0.6120$^{\dag}$ & 0.7835$^{\dag}$ & 0.3172$^{\dag}$ & - & - & - \\ 
APP+NoAspectFusion & 0.6117$^{\dag}$ & 0.7832$^{\dag}$ & 0.3188$^{\dag}$ & - & - & - \\    
    \bottomrule
  \end{tabular}
\vspace*{-2mm}
\end{table}

When we do not conduct aspect fusion and only use CLS as the final representation during fine-turning, the performance is also significantly better than BIBERT and competitive with the aspect fusion methods with ``CLS'' in it. It indicates that the aspect learning objectives (AP and APP) during pre-training are beneficial for the backbone model's parameters. It again confirms that introducing the special token ``OTHER'' during fine-tuning and training it from scratch is the reason that harms model performance. A side observation is that using the aspect presence prediction objective as an auxiliary pre-training task can improve some metrics a little, e.g., NDCG@10, showing that it does not have to be paired with the presence weighting mechanism when training MADRAL.  

\vspace*{-2mm}
\subsection{Comparisons of Aspect Representation}
We compare different ways of aspect representation in the model architecture in Table \ref{tab:exp_arep}. The reported numbers are based on their best fusion methods, i.e., presence weighting for extra k and CLS gating for the rest. Similar to Table \ref{tab:exp_fusion}, we show the performance of both using ``OTHER'' and ``CLS'' in aspect fusion. It is obvious that reusing the first k content tokens as aspect representations outperforms declaring extra k embeddings in the original MADRAL. Since reusing existing tokens will not face the issue of insufficient learning of brand-new parameters, it is not surprising that ``first k'' can be a better aspect representation option. Another interesting finding is that the gap between using ``CLS'' and ``OTHER'' for reusing content tokens is much smaller than ``Extra k''. This means that based on the pre-trained model that attaches the aspect learning with existing content tokens, it reduces the difficulty for the model to act effectively by learning an implicit embedding ``OTHER'' during fine-tuning.

\begin{table}
\vspace*{-2mm}
\setlength{\abovecaptionskip}{0cm}
\renewcommand{\arraystretch}{0.8}
\setlength{\belowcaptionskip}{-0.2cm}
\caption{Study of aspect representation approach. ``br'', ``co'', and ``ca'' are short for ``brand'', ``color'', and ``category''. The best results are in bold. $\dag$ and $\ddag$ indicates significant improvements over BIBERT and the ``Extra k" method, respectively.}
\centering
\label{tab:exp_arep}
  \begin{tabular}{lcccccc}
    \toprule
         \multirow{2}{*}{Method} & \multicolumn{3}{c}{``CLS'' in Fusion}&\multicolumn{3}{c}{``OTHER'' in Fusion} \\
     \cmidrule(lr){2-4} \cmidrule(lr){5-7}
 &R@100&R@500 & NDCG@10 & R@100 & R@500  & NDCG@10 \\
 \midrule
Extra k & 0.6148$^{\dag}$ & 0.7878$^{\dag}$ & 0.3184$^{\dag}$ & 0.5016 & 0.7121 & 0.2086 \\ 
\midrule
First k (br,co,ca) & 0.6216$^{\dag \ddag}$ & 0.7919$^{\dag \ddag}$ & 0.3285$^{\dag \ddag}$ & 0.6087$^{\ddag}$ & 0.7839$^{\dag \ddag}$ & 0.3157$^{\ddag}$ \\ 
First k (ca,co,br) & \textbf{0.6219}$^{\dag \ddag}$ &\textbf{0.7922}$^{\dag \ddag}$ & \textbf{0.3291}$^{\dag \ddag}$ & 0.6132$^{\dag \ddag}$ & 0.7859$^{\dag \ddag}$ & 0.3179$^{\dag \ddag}$ \\ 
Random k & 0.6188$^{\dag \ddag}$ & 0.7874$^{\dag}$ & 0.3230$^{\dag \ddag}$ & 0.6081$^{\ddag}$ & 0.7836$^{\dag \ddag}$ & 0.3113$^{\ddag}$ \\
    \bottomrule
  \end{tabular}
\vspace*{-5mm}
\end{table}

We also study whether the positions of the content tokens that the aspects are mapped to would affect model performance. 
By using the first 3 tokens to represent different aspects, i.e., (brand, color, category) and (category, color, brand), we do not see significant differences when CLS is in the fusion while the latter is better when ``OTHER'' is in the fusion. Since ``category'' is the most important aspect \cite{madr,sun2023pre}, it seems that mapping it to a higher position can help the model be stable when learning a new representation during fine-tuning. 
We also mapped the aspects to random k positions and conducted aspect learning, denoted as ``Random k''. As smaller positions are often more important in representing an item, we limited the random selection to within the first twenty positions. 
The performance of ``Random k'' (i.e., positions 3, 9, 15, and 7 as ``brand'', ``color'', ``category'', and ``OTHER'' when needed) is lower than ``First k” but still beats ``Extra k''. It implies that reusing positions at higher positions for aspect representation would be better, which is not surprising since the beginning tokens are usually more important than the others.

\begin{table}
\vspace*{-4mm}
\renewcommand{\arraystretch}{0.8}
\setlength{\belowcaptionskip}{-0.1cm}
\setlength{\abovecaptionskip}{0cm}
    \centering
    \caption{Accuracy@3 of the predicted values using the aspect embeddings learned after pre-training and fine-tuning.}
    \label{tab:aspect_acc}
    \begin{adjustbox}{width=0.9\textwidth}
    \begin{tabular}{ccccccc}
        \toprule
        \multirow{2}{*}{\textbf{Method}} & \multicolumn{2}{c}{\textbf{Category}} & \multicolumn{2}{c}{\textbf{Brand}} & \multicolumn{2}{c}{\textbf{Color}} \\
        \cmidrule(lr){2-3} \cmidrule(lr){4-5} \cmidrule(lr){6-7}
        & \textbf{pre-train} & \textbf{fine-tune} & \textbf{pre-train} & \textbf{fine-tune} & \textbf{pre-train} & \textbf{fine-tune} \\
        \midrule
        MTBERT(al=0) & 0.9728 & 0.1118 & 0.9727 & 0.0004 & 0.7843 & 0.0060 \\
        MTBERT(al=0.05) & 0.9728 & 0.4758 & 0.9727 & 0.0066 & 0.7843 & 0.2367 \\
        \midrule
        MADRAL-en-v1(al=0) & 0.9712 & 0.8786 & 0.9811 & 0.8251 & 0.7725 & 0.2008 \\
        MADRAL-en-v1(al=0.05) & 0.9712 & 0.9664 & 0.9811 & 0.9649 & 0.7725 & 0.7333 \\
        \bottomrule
    \end{tabular}
    \end{adjustbox}
\vspace*{-5mm}
\end{table}
\vspace*{-5mm}
\subsection{Effect of AL Coefficient $\lambda_f$ in Fine-Tuning}
\begin{figure}
\vspace*{-3mm}
\setlength{\belowcaptionskip}{-0.5cm}
\setlength{\abovecaptionskip}{-0.5cm}
    \centering
    \begin{minipage}{0.49\textwidth}
    \setlength{\belowcaptionskip}{0cm}
\setlength{\abovecaptionskip}{0cm}
        \includegraphics[width=0.9\textwidth]{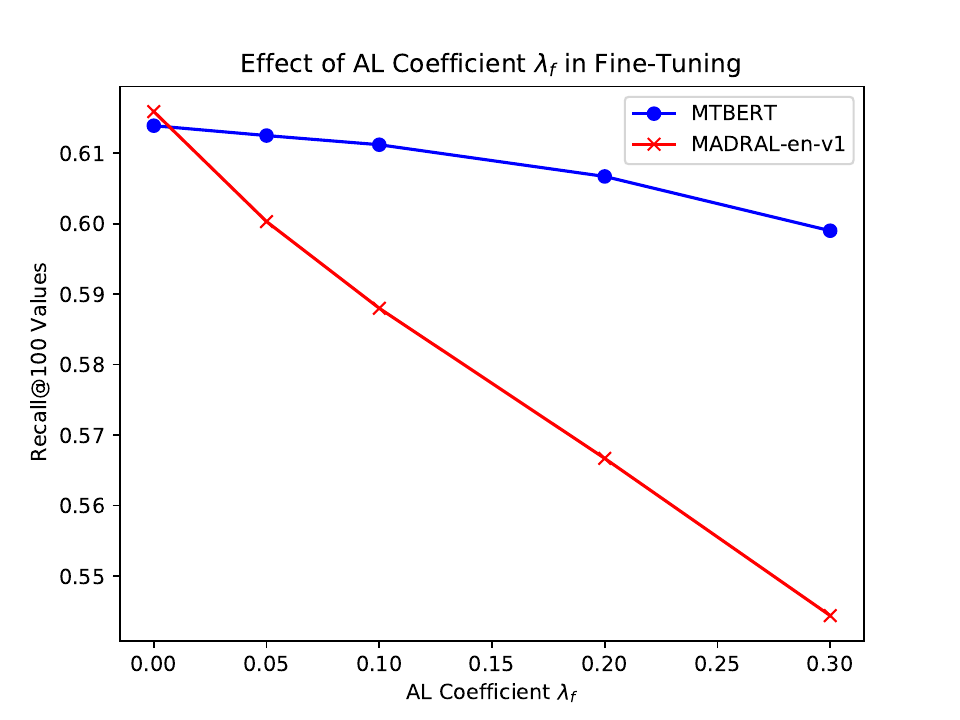}
        \caption{Effect of AL Coefficient $\lambda_f$}
        \label{fig:al_coe}
    \end{minipage}
    \begin{minipage}{0.49\textwidth}
    \setlength{\belowcaptionskip}{0cm}
\setlength{\abovecaptionskip}{0cm}
        \includegraphics[width=0.9\linewidth]{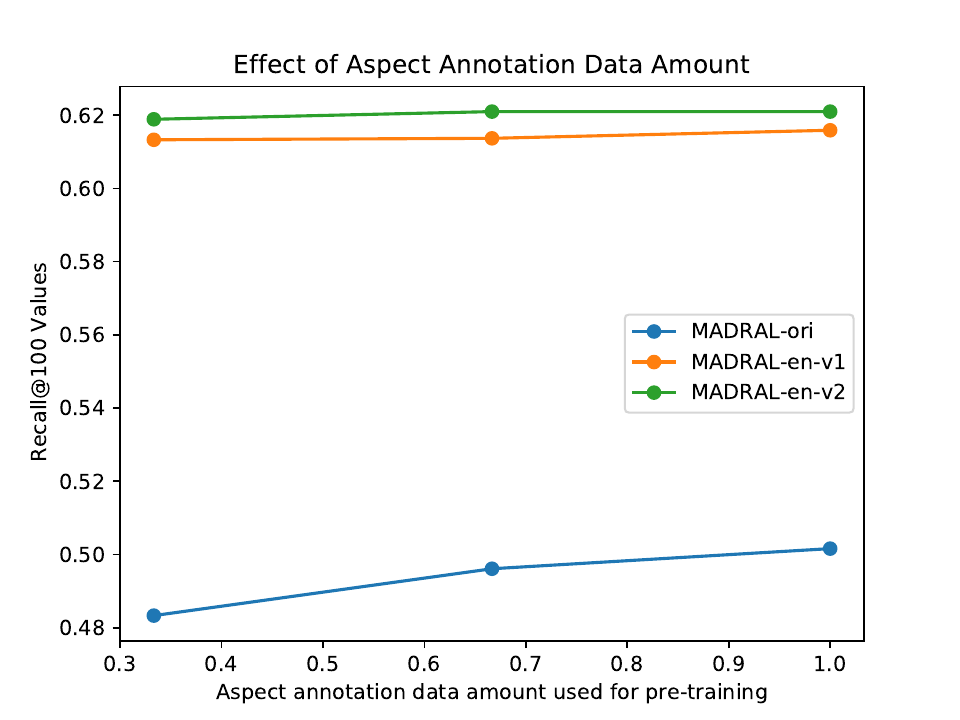}
        \caption{Effect of Annotation Amount}
        \label{fig:anno-amount}
    \end{minipage}
    \label{fig:hyper-effect}
\vspace*{-4mm}
\end{figure}

In the original paper \cite{madr}, the aspect learning objectives during fine-tuning are helpful for the retrieval performance but we find that they would harm both MTBERT and MADRAL. Figure \ref{fig:al_coe} and Table \ref{tab:aspect_acc} show the retrieval performance and the accuracy of aspect prediction when using different coefficients of aspect learning during fine-turning. From Table \ref{tab:aspect_acc}, we observe that a small amount of aspect prediction (AP) loss during fine-tuning will boost the AP accuracy. However, the one with higher AP accuracy has worse retrieval performance, as shown in Figure \ref{fig:al_coe}. The more AP is used, the more retrieval performance will drop. This implies that the learning objectives between relevance matching and aspect prediction guide the model in different directions.  

\vspace*{-10mm}
\subsection{Effect of Aspect Annotation Amount}
We divide the aspect annotations of the item into three partitions and gradually include more aspect annotations for aspect learning during pre-training. The MLM loss on the entire corpus is always used for pre-training. Figure \ref{fig:anno-amount} illustrates the effect of aspect annotation amount on the original MADRAL and our two versions of enhancements. For MADRAL-ori, more annotations help the model perform better (i.e., from 0.4833 to 0.5016), which is consistent with our speculation that the extra k aspect embeddings require more aspect annotations for sufficient learning. When ``CLS'' replaces ``OTHER'' during fusion (in MADRAL-en-v1), more annotations bring fewer benefits (i.e., from 0.6133 to 0.6159), indicating that this fusion manner requires fewer aspect annotations to act effectively. When the first k content tokens are adjusted by aspect learning (in MADRAL-en-v2), the recall at 100 saturates with 2/3 aspect annotations. It shows that when aspect learning is used for refining existing important content tokens (first k), even fewer aspect annotations are needed to act effectively.

\vspace*{-3mm}
\section{Conclusion}
In conclusion, this paper presents a critical examination of the first multi-aspect dense retrieval model, MADRAL. Observing its failure on the public MA-Amazon data, we conduct a thorough investigation into MADRAL's components of aspect representation and fusion. We propose several alternative approaches for each component and compare them with their original counterparts.
We find that it has a detrimental effect on retrieval performance to learn implicit semantics with the special aspect ``OTHER''. In contrast, the proposed variants, including replacing ``OTHER'' with "CLS" (that represents the overall content semantics explicitly) and representing aspects with the first few content tokens, have demonstrated significant improvements in retrieval performance.
\vspace*{-2mm}
\section*{Acknowledgments}
This work was funded by the National Natural Science Foundation of China (NSFC) under Grants No. 62302486, the Innovation Project of ICT CAS under Grants No. E361140, the CAS Special Research Assistant Funding Project, the Lenovo-CAS Joint Lab Youth Scientist Project, and the project under Grants No. JCKY2022130C039. 

%
%
%
%

\bibliographystyle{splncs04}
\bibliography{reference}

\end{document}